\documentclass{article}

\usepackage{arxiv}

\usepackage{amsmath,amsfonts,amssymb, amsthm}

\usepackage[dvipsnames]{xcolor}
\definecolor{niceRed}{RGB}{190,38,38}
\definecolor{niceBlue}{HTML}{0466a7}

\usepackage[utf8]{inputenc} 
\usepackage[T1]{fontenc}    
\usepackage{hyperref}       
\hypersetup{
	colorlinks = true,
	urlcolor = {black},
	linkcolor = {niceBlue},
	citecolor = {niceRed} 
}

\usepackage{natbib}
\bibpunct[, ]{(}{)}{,}{a}{}{,}%
\def\BIBand{and}%

\usepackage{xurl}           
\usepackage{booktabs}       
\usepackage{multirow}
\usepackage{nicefrac}       
\usepackage{microtype}      
\usepackage{cleveref}       
\usepackage{graphicx}
\usepackage{doi}
\usepackage{tikz}
\usepackage{mathtools}
\usepackage{pifont} 		

\usepackage[labelfont=sf]{subcaption}
\usepackage{booktabs}	
\usepackage{multirow}
\usepackage{algpseudocode}
\usepackage[ruled]{algorithm2e}
\usepackage{float}

\usepackage{mathtools}

\newtheoremstyle{spaced}%
  {10pt}   
  {10pt}   
  {\itshape} 
  {}       
  {\bfseries} 
  {.}      
  { }      
  {\thmname{#1}\thmnumber{ #2}\thmnote{ (#3)}} 
\makeatother

\theoremstyle{spaced}

\definecolor{lightgray}{gray}{0.7}
\definecolor{synblue}{RGB}{0,66,118}
\definecolor{synred}{RGB}{143,53,71}
\definecolor{syngreen}{RGB}{34,143,156}
\definecolor{darkviolet}{rgb}{0.58, 0.0, 0.83}

\definecolor{customblue}{HTML}{4C72B0}
\definecolor{customred}{HTML}{C44E52}

\makeatletter

\newcommand{\Acal}{\mathcal{A}}

\newcommand{\Ncal}{\mathcal{N}}
\newcommand{\Scal}{\mathcal{S}}

\DeclarePairedDelimiter{\abs}{\lvert}{\rvert}

\newcommand{\mi}{{-i}} 

\title{Algorithmic Pricing and Algorithmic Collusion}

\newif\ifuniqueAffiliation
\uniqueAffiliationtrue

\author{
	Martin Bichler \\
	\small Department of Computer Science\\
	\small Technical University of Munich\\
	\small \texttt{m.bichler@tum.de} \\
	\And
	Julius Durmann\\
	\small Department of Computer Science\\
	\small Technical University of Munich \\
	\small \texttt{julius.durmann@tum.de}\\
	\And
	Matthias Oberlechner\\
	\small Department of Computer Science\\
	\small Technical University of Munich \\
	\small \texttt{matthias.oberlechner@tum.de}\\
}

\renewcommand{\headeright}{}
\renewcommand{\undertitle}{}
\renewcommand{\shorttitle}{}

\hypersetup{
	pdftitle={Algorithmic Pricing and Algorithmic Collusion},
	pdfsubject={cs.GT},
	pdfauthor={Bichler et al.},
	pdfkeywords={} ,
}

\begin{document}
\maketitle

\begin{abstract}
	The rise of algorithmic pricing in online retail platforms has attracted significant interest in how autonomous software agents interact under competition.
	This article explores the potential emergence of algorithmic collusion — supra-competitive pricing outcomes that arise without explicit agreements — as a consequence of repeated interactions between learning agents.
	Most of the literature focuses on oligopoly pricing environments modeled as repeated Bertrand competitions, where firms use online learning algorithms to adapt prices over time. While experimental research has demonstrated that specific reinforcement learning algorithms can learn to maintain prices above competitive equilibrium levels in simulated environments, theoretical understanding of when and why such outcomes occur remains limited. This work highlights the interdisciplinary nature of this challenge, which connects computer science concepts of online learning with game-theoretical literature on equilibrium learning.
	We examine implications for the Business \& Information Systems Engineering (BISE) community and identify specific research opportunities to address challenges of algorithmic competition in digital marketplaces.
\end{abstract}


\emph{\textbf{Note\,}} This article has been accepted in \emph{Business \& Information Systems Engineering (BISE)}.

\section{Introduction}

Worldwide, companies are increasingly using algorithms and artificial intelligence (AI) in order to power their operations, from product development to manufacturing and marketing, and product pricing is no exception. This new type of information systems based on learning algorithms has drawn considerable attention aiming to understand their societal effects \citep{lysyakov2023threatened, lu20241+, fugener2021will}. An important question in this overall stream is how learning algorithms are used for pricing \citep{brackmann2024art}. 
Algorithmic pricing is a practice where software agents automatically determine prices for items for sale, in order to maximize the seller's profits. This practice is increasingly common in online retail markets. \citet{chen2016empirical} estimated that by 2015, algorithms were used in setting prices for roughly one-third of the top 1600 products on Amazon. By 2018, the average product price on Amazon reportedly changed every ten minutes and adapts to market conditions.\footnote{\url{https://www.businessinsider.com/amazon-price-changes-2018-8}} Since then, an industry has developed around automated pricing software.

The prices determined by the different firms depend on each other's actions, and the overall environment in which they operate is that of an oligopoly competition. Economic theory has long aimed to predict the outcome of such competitive situations. Models of oligopoly competition assume some knowledge about the demand of customers, such as a readily available demand function, and then they determine a Nash equilibrium price. 
For example, in the celebrated Bertrand competition model, companies producing identical (homogeneous) products simultaneously choose their pricing strategies based on a given demand function that takes the prices of the competitors into account \citep{bertrand1883book}. 
In this model, the competitors play their equilibrium strategy from the start. 
Usually, sellers entering the market have limited information about competitors' costs or customer demand at different prices, and they must learn over time which prices optimize profit.
Additionally, changes in demand and supply over time require recalculating the equilibrium strategy, which emphasizes the need for an adaptive and learning pricing software agent.
From the perspective of an individual seller, algorithmic pricing aims to solve an online learning problem \citep{shalev-shwartzOnlineLearningOnline2011}, where the seller's actions are the prices they set, and the objective is to maximize profit. But what is the outcome of markets with such learning agents? Can we assume that they converge to an equilibrium price? 

Recent experimental research showed that learning algorithms can lead to prices higher than the Nash equilibrium in the static oligopoly pricing game \citep{waltman2008q, calvano_algorithmic_2019, abada_artificial_2023, klein_autonomous_2021, abada_collusion_2024, brown_competition_2023}.
This phenomenon is referred to as \textit{algorithmic collusion}. 
Explicit {collusion} refers to anti-competitive conducts that are maintained with explicit agreements. 
Firms interact directly and agree on the optimal level of price or output \citep{oecd2017}. In contrast, ``tacit collusion refers to forms of anti-competitive co-ordination which can be achieved without any need for an explicit agreement, but which competitors are able to maintain by recognizing their mutual interdependence. In a tacitly collusive context, the non-competitive outcome is achieved by each participant deciding its own profit-maximizing strategy independently of its competitors.'' \citep{oecd2017}  
\textit{Algorithmic collusion} is a form of tacit collusion where supra-competitive outcomes different from the Nash equilibrium of the static game-theoretical model of competition are produced by learning algorithms without being programmed to produce those outcomes. Similar definitions are provided by \citet{den2023mathematical} and \citet{abada2024algorithmic}.

There are also cases suggesting that algorithmic collusion happens in real-world markets. For example, \citet{assad2020algorithmic} showed that margins increased 28\% in local duopoly retail gasoline markets in Germany when both firms adopted algorithmic pricing software, while there was no price change in local monopolies. The paper finds pricing algorithms can learn tacitly collusive pricing strategies which are legal in most jurisdictions without explicit communication. An investigation in the UK revealed that online poster retailers were using simple pricing algorithms in the context of a horizontal cartel among retailers to coordinate their prices on Amazon.\footnote{\url{https://www.gov.uk/cma-cases/online-sales-of-discretionary-consumer-products}} 
The phenomenon raised concerns among regulators as it may reduce consumer welfare. In recent years, several competition authorities, including France and Germany, Denmark, Japan, Norway, and Sweden, have published policy papers considering the relationship between algorithms and competition \citep{oecd2024}. 

However, the magnitude of the threat from algorithmic collusion by autonomous self-learning algorithms in other markets is still disputed in the academic literature. Ultimately, algorithmic collusion touches on a deep and non-trivial question, that of learning in games. Under which conditions do learning algorithms converge to an equilibrium in repeated play and when is this not the case? When do we see algorithmic collusion, inefficient price cycling, or even chaotic price dynamics as a result of automated pricing decisions? Currently, there is no comprehensive theory that would provide answers to these questions. Up to now, there are only very few articles in Business \& Information Systems Engineering outlets on algorithmic collusion \citep{kang2022raising,douglas2024naive,deng2024algorithmic,bichler2023soda}. The topic is central for the welfare of electronic markets today and it draws on learning algorithms from computer science as well as game-theoretical models as they have been developed in economics. 

The aim of this catchword article is to discuss algorithmic collusion and make the basic problem and the related questions accessible to a broader community. This will allow us to point to avenues for future research in Business \& Information Systems Engineering.

\section{Oligopoly Pricing}

Pricing in online retail markets is an environment that can be modeled as a Bertrand competition \citep{bertrand1883book}. In this model, firms compete by setting prices for a homogeneous product. The demand for the product depends on the prices set by all firms. The firms' objective is to maximize their profit, which is the revenue from selling the product minus the cost of producing it. The revenue depends on the price set by the firm and the demand for the product. The demand is a function of the prices set by all firms. The cost of producing the product is typically assumed to be linear in the quantity produced.

More formally, the Bertrand competition can be described as a \textit{normal-form game}, where $n$ players (or firms) make decisions at the same time.
Each player $i = 1, 2, \dots, n$ has a set of possible actions $a_i \in \Acal_i$ to choose from - like different prices they could set for their products. 
Each player's goal is to maximize their payoff they get based on everyones actions, which is given by their individual payoff function $u_i: \Acal_1 \times \dots \times \Acal_n \rightarrow \mathbb{R}$.
Normal-form games are used to model environments where multiple agents interact and influence each others' outcomes. Other examples include, for instance, the Cournot oligopoly, platform competition, auctions, and contests. Most of the literature on algorithmic collusion is based on the classic Bertrand competition model. 

In a Bertrand competition, the firms' action  $a_i \in \Acal_i$ is the price for a good it wants to produce. All firms produce the same good, and they all take action simultaneously. 
Firms compete for demand with their prices and affect each others' revenues. The demand $d_i(a_i,a_{-i})$ depends on all actions, their own actions $a_i$ and the others' actions $a_{-i}$, and is decreasing in their own price. 
Assuming that the cost functions are linear in the demand, the payoff (or utility) function of firm $i$ can be described by
\begin{equation}
    u_i(a_i, a_\mi) =  d_i(a) \cdot (a_i - c_i).
\end{equation}

This model allows for various assumptions about consumer demand (e.g., all-or-nothing or logit), and the equilibrium prices in these scenarios have been analyzed. 
In the standard case, with \textit{all-or-nothing demand}, the firm with the lowest price gets all the demand. If multiple firms offer the lowest price, the demand is shared equally. This leads to a demand function given by
\begin{equation} \label{eq:demand_standard}
    d_i(a_i, a_{-i}) = \begin{cases} \frac{D}{n_{min}} (1- a_i) &\text{if } i \in \arg \min_{j \in \Ncal} a_j \\ 0 &\text{else} \end{cases}, 
    \tag{all-or-nothing} 
\end{equation}
where $D > 0$ is the maximum total demand, $\Ncal$ is the set of market participants, and $n_{min} := \abs{\arg \min_{j \in \Ncal} a_j}$ is the number of firms with the lowest price.

Another demand model used in a widely cited article by \citet{Calvano2020}, is the \textit{multinomial or logit demand}. In this setting, the demand is split between the $n$ agents/goods and some outside good (indexed with $0$) according to
\begin{equation}\label{eq:logit_demand}
     d_i(a_i,a_\mi) = \dfrac{\exp \left( \frac{\alpha_i - a_i}{\mu} \right) }{ \exp \left( \frac{\alpha_0}{\mu} \right) + \sum_{j=1}^n \exp \left( \frac{\alpha_j - a_j}{\mu} \right)}. 
     \tag{logit}
\end{equation}
The parameters $\alpha_i > 0$ capture different product quality indices and $\mu > 0$ models a product differentiation between the goods, i.e., if $\mu \rightarrow 0$, the goods are perfect substitutes. 

In a standard Bertrand model with homogeneous products and symmetric firms, prices equal marginal costs in equilibrium. However, this changes with asymmetries or product differentiation. 
With asymmetric costs between firms, the Nash equilibrium typically involves the low-cost firm pricing at or just below the marginal cost of the high-cost firm. With product differentiation (as in logit demand), equilibrium prices are typically above marginal costs due to firms having some market power. These equilibrium prices are below a monopoly price because the competition drives down profits.

Note that the static model of Bertrand competition has been extended to repeated games. A repeated game consists of a number of repetitions of some base game (called a stage game). 
In such dynamic settings, a price above the Nash equilibrium of the stage game can be an equilibrium of this repeated game depending on the strategy used by each firm and other exogenous factors \citep{maskin1988theory}. 
Actually, the Folk Theorem in repeated game theory shows that if players are sufficiently patient, a wide range of payoffs can be sustained as Nash equilibria, including outcomes that are better for each player than their min-max (or punishment) payoff \citep{maschler2020game}.
The literature on algorithmic collusion takes the Nash equilibrium of the stage game of such repeated games as a baseline to determine algorithmic collusion \citep{abada2024algorithmic}. 

Much of the discussion around algorithmic collusion relies on simulations of learning algorithms in repeated Bertrand competition models with fixed sellers and specific demand functions. Although real-world scenarios may be more complex, this abstraction allows for an analysis of the model and the computation of its static Nash equilibrium. This knowledge enables comparative statics based on cost and demand knowledge, which is typically unavailable in empirical studies. If algorithmic collusion does not arise in a repeated Bertrand pricing game, it may be even less likely in more complex scenarios with fluctuating demand and supply. If it does arise, it is a strong indicator that the phenomenon is not limited to the specific model but may also occur in more complex settings.


In what follows, we introduce important online learning algorithms, the emerging literature on algorithmic collusion, and the key results from the literature on learning in games that are related to it. These literature streams are important for the study of algorithmic collusion and inter-related.

\subsection{Online Learning}

A major challenge in developing algorithmic pricing agents is deciding whether to focus on short-term profit (by exploiting a known high-yield price) or on exploring alternative prices that may lead to better long-term outcomes. Online learning algorithms are designed to balance this exploration-exploitation trade-off and can handle large product portfolios effectively \citep{bubeck2011introduction}. On online platforms, these algorithms operate with bandit feedback, meaning that after setting a price, a seller observes the profit associated with that specific price. The multi-armed bandit model is especially relevant to algorithmic pricing, a connection recognized early on. Bandit algorithms were proposed for pricing as far back as \citet{rothschild1974two}, well before digital marketplaces emerged. Today, multi-armed bandit algorithms for pricing are widely studied in academia \citep{trovo2015multi, den2015dynamic, bauer2018optimal, mueller2019low, elreedy2021novel, taywade_multi-armed_2023, qu24, kasa2021dependency}, and practitioners also provide numerous resources on implementing these algorithms.\footnote{\url{https://towardsdatascience.com/dynamic-pricing-with-multi-armed-bandit-learning-by-doing-3e4550ed02ac}, \url{https://www.griddynamics.com/blog/dynamic-pricing-algorithms}}

Online optimization and learning algorithms are prime candidates for pricing algorithms on online platforms \citep{mueller2019low, elreedy2021novel, taywade_multi-armed_2023, qu24}. 
Online optimization is concerned with making sequential decisions in an unknown environment with the goal of optimizing a performance metric over time. 
Let us briefly introduce the basic model of online learning in the context of algorithmic pricing in its most abstract form (see Algorithm 1). At each stage $t = 1,2,\dots$, the agent chooses an action $a_t \in \Acal $ from some action set according to some strategy $s_t \in \Scal$ and gets a payoff $u_t(a_t)$. In algorithmic pricing, this action would be the price of an agent. The agent would leverage the information about the utility, i.e., feedback, she gets in order to update his actions or prices.

\begin{algorithm}
\caption{Online Learning}
\SetKwInOut{Require}{Require}
\Require{action set $\mathcal{A}$, sequence of payoff functions $u_t: \mathcal{A} \to \mathbb{R}$}
\For{$t = 1, 2, \dots$}{
    select action $a_t \in \mathcal{A}$ according to strategy $s_t$\; 
    realize payoff $u_t(a_t)$ and observe feedback $f_t$\;
    update strategy $s_t \leftarrow s_{t+1}$ using feedback $f_t$\;
}
\end{algorithm}

Algorithms are widely analyzed in two models, the adversarial model and the stochastic model. In both models, the objective goal is to maximize the expected reward by selecting the best action(s). In the stochastic setting, we assume that the rewards are drawn independently and identically from some underlying distribution that does not change over time. In the adversarial model, the input can be chosen by an adversary that can react to the agent's past decisions and its algorithm. There are different types of feedback available to the agents in online optimization algorithms such as bandit or gradient feedback. 

Real-world implementations of learning pricing agents are best modeled with bandit feedback. This means that the feedback consists of a pointwise evaluation of the payoff function at the chosen action: $f_t \,\hat{=}\, u_t(a_t)$. For example, after setting a price in an oligopoly competition, an agent observes a specific profit in the next period. They often have no or only incomplete information about the demand model or the strategies used by all other sellers. This is particularly true on large online retail platforms where there are many substitutes for a good. 

In the stochastic or adversarial model, one can analyze the performance of an algorithm. The key characeristic in this literature is regret, i.e. the difference between the cumulative payoff if the algorithm played the best fixed price in hindsight and the cumulative payoff of the learning algorithm. For some online learning algorithms the  regret vanishes over time. These algorithms are also referred to as no regret algorithms. 
As an example, Exponential Weights (or the Exp3 variant) is a well-known online learning algorithm that uses bandit feedback and updates the weights associated with each action based on the cumulative reward observed for that action. The algorithm then chooses actions with probabilities proportional to their weights. Exp3 is a no regret algorithm in the adversarial model. 

There are also academic articles using (deep) reinforcement learning algorithms for algorithmic pricing \citep{rana2014real, kastius2022dynamic, deng2024algorithmic}. Different from the online learning algorithms discussed so far, reinforcement learning algorithms allow the agent to take into account the state of a system. This state could be historical prices, the day of the week, or other variables that are potentially relevant. With an increasing state and action space, learning requires a lot of training rounds and doesn't scale well. 
Interestingly, the most prevalent reinforcement learning algorithm in the algorithmic collusion literature is Q-learning. However, in related articles, the state space is limited to the current price, letting the algorithm resemble an online learning algorithm. Yet, \citet{lambin2024less} shows that algorithmic collusion can arise with Q-learning even in a stateless version of Q-learning. 

\subsection{Algorithmic Collusion}

\begin{figure}[h]
    \begin{center}
    \begin{subfigure}{.4\textwidth}
        \centering
        \includegraphics[width=0.8\linewidth]{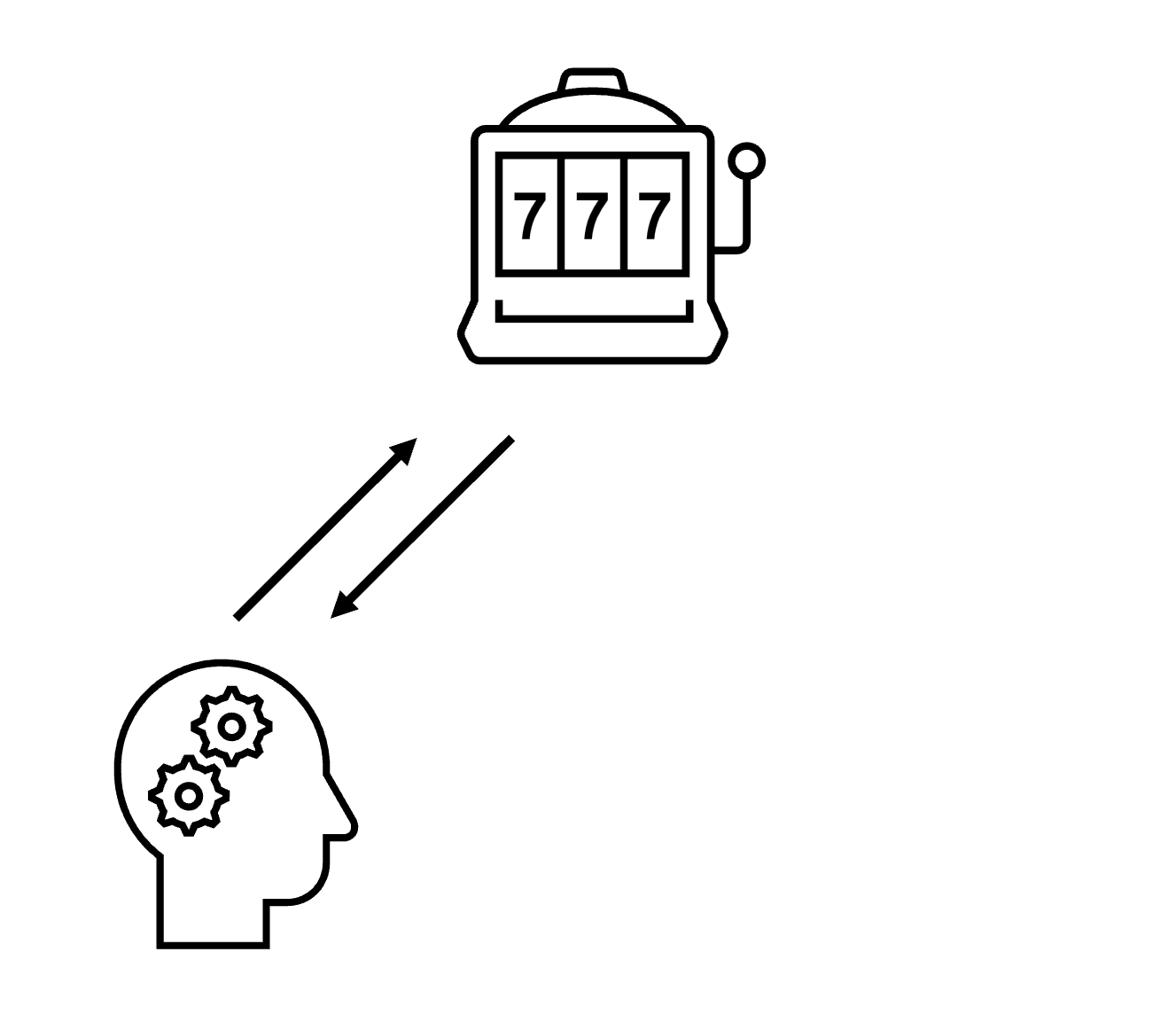}
        \caption{Online Learning}\label{fig:onl_learning}
    \end{subfigure}%
    \begin{subfigure}{.4\textwidth}
        \centering
        \includegraphics[width=0.8\linewidth]{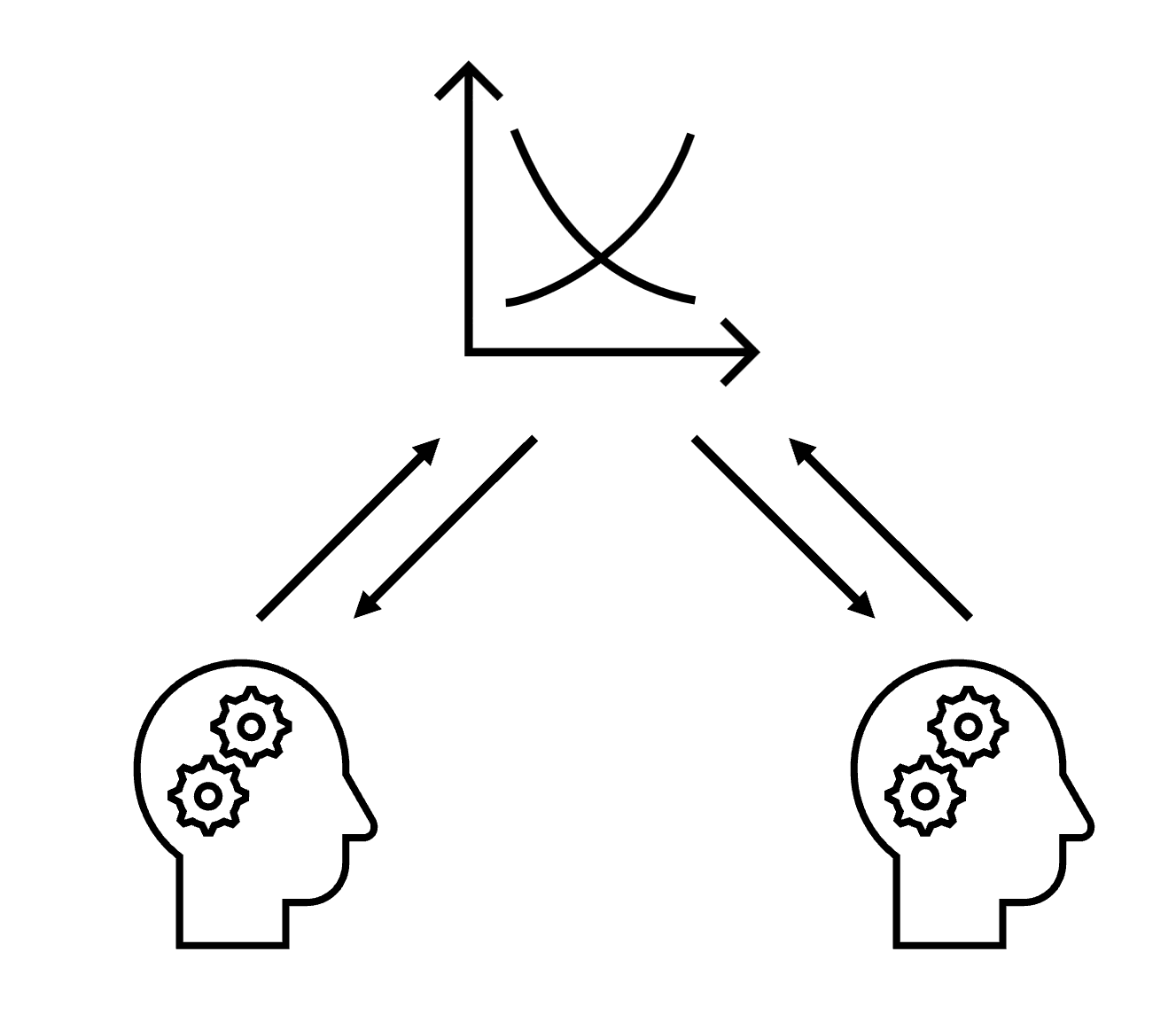}
        \caption{Equilibrium Learning}\label{fig:equil_learning}
    \end{subfigure}
    \caption{Learning Agents in Different Contexts}
    \label{fig:learning_agents}
    \end{center}
    \small 
    In the classical online learning setting, a single agent selects actions and observes stochastic (or adversarial) payoffs. By contrast, algorithmic collusion studies the outcomes when multiple agents interact using online learning algorithms. Understanding the results of these multi-agent learning processes requires consideration of the algorithm, but also of the structure and properties of the underlying game, which is analyzed in equilibrium learning.
\end{figure}

While the literature on online learning gives us methods to find optimal strategies in stochastic or even adversarial settings, it does not capture the interaction of multiple agents using such algorithms (see Figure \ref{fig:learning_agents}). 
The literature on algorithmic collusion attempts to fill this gap and has attracted considerable attention from the research community and policymakers.

Most of this literature analyzes specific algorithms such as Q-learning for specific model variations, i.e., Bertrand oligopolies with standard all-or-nothing demand, linear, or logit demand. 
\citet{Calvano2020} analyzed a Bertrand competition with logit demand and constant marginal cost. They found that when all agents employ Q-learning, the competition of these agents can lead to supra-competitive prices higher than the Nash equilibrium.  
A related sequential move pricing duopoly environment with linear demand (instead of the simultaneous move Bertrand model in \citet{Calvano2020}) was analyzed by \citet{klein_autonomous_2021}, who also found collusion with Q-learning agents. 
\citet{asker2022artificial} detected in their experiments on Bertrand competition with standard (all-or-nothing) demand that collusion depends on specifics of the Q-learning algorithm (e.g., synchronous vs. asynchronous updating).

In contrast, \citet{abada_collusion_2024} analyzed Q-learning in Bertrand oligopolies and showed that Q-learning algorithms with sufficiently large $\epsilon$-greedy exploration exhibit no collusion. \citet{den_boer_artificial_2022} provided a detailed analysis of the inner workings of Q-learning and argue that Q-learning would not lead to collusion easily. 
In addition, \citet{eschenbaum2022robust} criticized the claim that algorithms can be trained offline to successfully collude online in different market environments. The authors found that collusion breaks down when collusive reinforcement learning policies are extrapolated from a training environment to the market. 
While most of this experimental literature on algorithmic collusion is based on Q-learning, there is little evidence that this algorithm is particularly important or widespread for algorithmic pricing.  
More recently, \citet{hansen_frontiers_2021} analyzed the price levels that arise in a duopoly setting where agents based on the UCB ("Upper Confidence Bound") algorithm determine prices. They ran a series of experiments where a variant of symmetric UCB algorithms interact simultaneously in a Bertrand economy competition with linear demand functions. The agents observed a perturbed estimate of their revenues which are a result of their prices and the corresponding demand. \cite{hansen_frontiers_2021} found that sometimes agents explore prices in a correlated manner, giving rise to supra-competitive outcomes. 

The literature on algorithmic collusion in Information Systems is still scarce. \citet{kang2022raising} and \citet{douglas2024naive} analyze the possibility of collusion in a repeated Prisonner's Dilemmata. \citet{deng2024algorithmic} discuss deep reinforcement learning in a repeated Bertrand competition, while \citet{bichler2023low} analyze the phenomenon in the context of display advertising auctions. 

Online learning algorithms address problems where agents optimize against an unknown and independent stochastic process. However, game-theoretical problems such as the Bertrand competition differ because each player's actions impact the objectives of others. In games, the Nash equilibrium (NE) represents a situation where no agent has an incentive to unilaterally deviate. Note that this is different from the adversarial setting in optimization discussed earlier, because each agent aims to maximize his payoff. If agents are not in equilibrium, then individual agents have an incentive to deviate from the current strategy profile, and the outcome is not stable. Independently, one might ask if the equilibrium reached maximizes welfare.

\subsection{Learning in Games} \label{sec:learning_games}

Although, the term algorithmic collusion is relatively new, the topic is related to a long standing stream of literature in game theory. Actually, the question if learning agents converge to a Nash equilibrium in repeated play is as old as the concept of the Nash equilibrium itself \citep{brown1951IterativeSolutionGames}. Actually, even Cournot's study of duopoly competition via quantity \citep{cournot1838recherches} already introduced a particular learning process. However, when learning algorithms converge to the Nash equilibrium and which properties they need to possess, is still not fully resolved \citep{young2004strategic, cesa2006prediction}.

Research on learning in games \citep{young2004strategic, foster1997CalibratedLearningCorrelated} has shown that not all games can be learned \citep{hart2006stochastic, milionis2022nash}: learning dynamics may cycle, diverge, or be chaotic \citep{mertikopoulos2018cycles, bailey2018multiplicative}. 
While there is no comprehensive characterization of games that are ``learnable'', there are some important results regarding learners. A classical result is that the class of no-regret learning algorithms converges to the so-called \emph{coarse correlated equilibrium} (CCE) of a game \cite{fudenberg1999TheoryLearningGames}. 
CCEs are superclasses of Nash equilibria. However, CCEs can also contain dominated strategies and the set of CCEs in a game can be very large. For the analysis of algorithmic collusion, we want to understand whether learning algorithms converge to a Nash equilibrium. 

Less is known about conditions of games in which learning algorithms converge to a Nash equilibrium. \citet{monderer1996potential} introduced the class of \textit{potential games}, and they showed that Cournot oligopolies with linear price or cost functions are potential games.  
Potential games are guaranteed to have at least one pure Nash equilibrium. Importantly, it was shown that several bandit algorithms converge to a Nash equilibrium in potential games \citep{palaiopanos2017multiplicative, cohen2017learning}. However, while Cournot oligopoly models are potential games, this property rarely holds in other economic games.

Another central condition for which positive results are known is that of the \textit{(payoff) monotonicity} of a game. Games that admit a strictly concave potential are strictly monotone \citep{mertikopoulos2019learning}. 
Games that satisfy this condition have a unique Nash equilibrium.
It is known that simple algorithms such as projected gradient ascent converge to an equilibrium of strictly monotone games \citep{dong2018inertial}. Yet, the class of games that are strictly monotone is rather restricted. 

Overall, the class of games for which learning algorithms converge to a Nash equilibrium is not well understood. More specifically, not much is known about properties of Bertrand competition models that would guarantee convergence to a Nash equilibrium. Note that each demand model assumed in the Bertrand competition model leads to a different game and thus might have different convergence properties.

\section{Implications for BISE Research}

A lot of the research in BISE and more broadly in the economic sciences aims to understand human behavior in certain market interactions. Electronic markets have been a central research topic in BISE for many years \citep{malone1987electronic, schmid2000elektronische, bichler2010research}. 
More and more markets are automated with learning agents, and we need to understand if these markets are in equilibrium and if they lead to efficient outcomes. These questions are not new, but the presence of learning algorithms is. 

\begin{figure}[h]
    \begin{center}
        \includegraphics[width=0.9\linewidth]{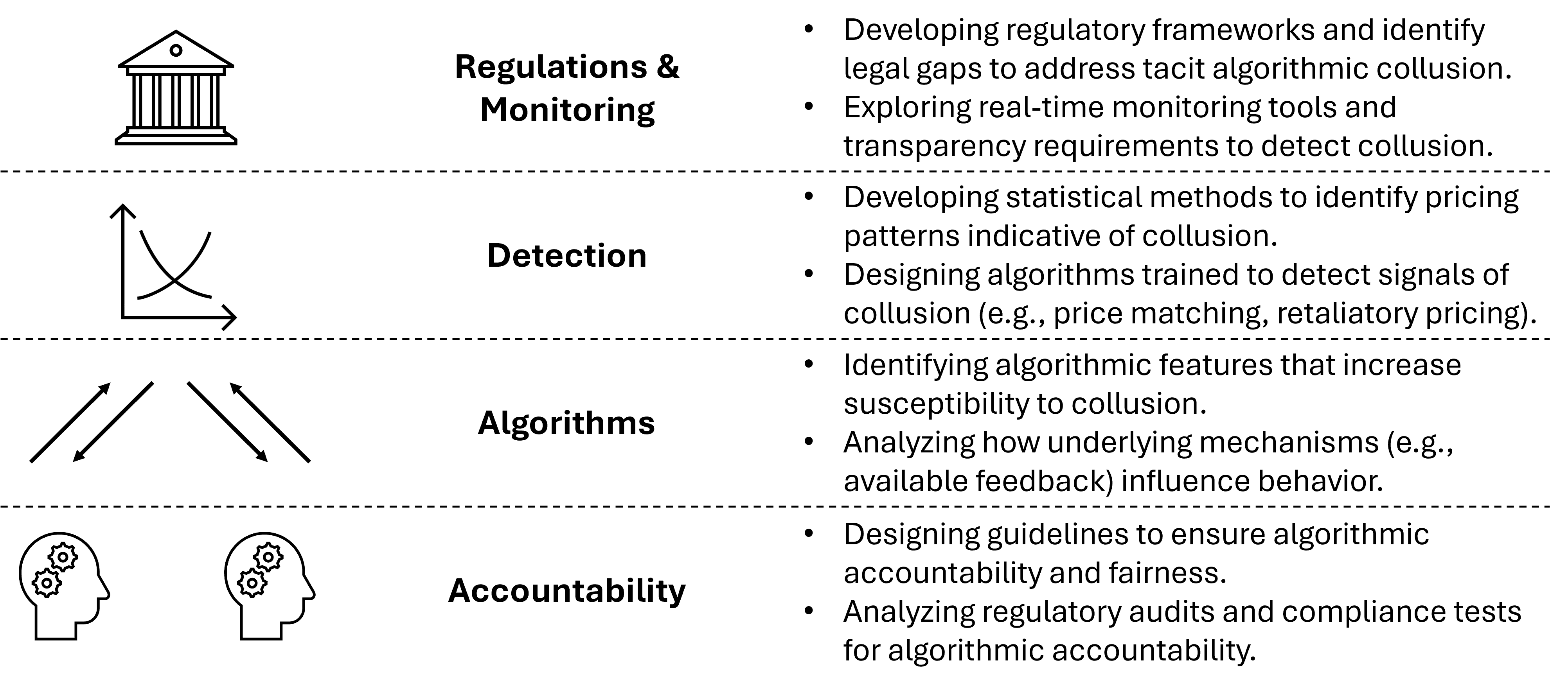}
        \caption{Overview of BISE Research Opportunities}\label{fig:onl_learning}
        \label{fig:bise_research}
    \end{center}
    \small
\end{figure}

Pricing agents on retail platforms such as Amazon are an example, and so are display ad auctions. The question of how these agents interact and what outcomes they produce is of great interest to researchers and policymakers. Does the use of learning algorithms in pricing lead to efficient equilibrium outcomes or does it jeopardize consumer welfare by leading to algorithmic collusion?
Algorithmic collusion challenges traditional theories that assume firms cannot sustain collusive arrangements without explicit coordination. 
The question has implications on policy, competition law, and the development of responsible algorithmic practices in pricing. In what follows, we discuss a variety of research questions that the BISE community is well-equipped to address. We provide a short overview in Figure~\ref{fig:bise_research}.

\paragraph{Algorithms}
A key question in the study of algorithmic collusion is determining which types of algorithms are more prone to collusive behavior. This inquiry involves examining specific properties of pricing algorithms that may influence their tendency to collude. Additionally, the assumptions within game-theoretical models can affect how different algorithms converge to equilibrium: some model setups may facilitate collusive outcomes for certain algorithms, while others may not. Although there are initial findings on Q-learning and some bandit algorithms within specific Bertrand competition models, a comprehensive understanding of these dynamics constitutes a wide open research question.

The type of feedback an algorithm receives significantly impacts its potential for collusion. Algorithms with bandit feedback, receiving only information on the outcomes of their own actions, are likely to behave differently than those which can access more information about the environment or competitors' actions. State-based information, like historical prices or observed demand patterns, can further enhance an algorithm's capacity to predict optimal prices, potentially leading to tacit collusion. Research is needed examining how different levels of information influence the emergence of collusive strategies. 

\paragraph{Detection}
Detecting algorithmic collusion presents a major challenge for regulators because price patterns resulting from collusion can closely resemble those from equilibrium strategies, especially in dynamic markets. Statistical methods might help to identify pricing patterns that may indicate collusion \citep{bonjour2022information}. Another area of focus is the development of algorithms trained to recognize subtle signals of collusion, such as price matching or retaliatory price adjustments \citep{xu2024learning}. Effective detection methods could play a key role in helping regulators monitor and address collusive pricing practices on digital platforms.

\paragraph{Regulations \& Monitoring}
Even if a competition authority were to identify a potential case of tacit collusion, the current state of the law could make such practice irreproachable in the absence of explicit communication or contact among the companies using such autonomous algorithms. 
Existing competition laws often focus on explicit, human-driven collusion rather than implicit algorithmic cooperation. New regulatory frameworks may be necessary to address algorithmic behaviors that lead to collusive outcomes, even without direct communication between firms.
The OECD and other regulators are aware of shortcomings of the existing legislation, and several calls to action have been made for policy changes to address this potential enforcement gap \citep{oecd2024}. For example, the European Commission's revised guidelines on horizontal cooperation agreements, adopted in 2023, stipulate that an explicit agreement among competitors to use the same pricing algorithm is considered an infringement of article 101 of the Treaty on the Functioning of the European Union.
Note that this action addresses a different form of collusion where several firms use the same third-party pricing software to determine their prices. This may result in a hub-and-spoke situation that can facilitate information exchanges in the context of an agreement or concerted practice \citep{oecd2024}. This is different from the type of algorithmic collusion we discuss in this paper, where firms use their own pricing algorithms and learn to collude without explicit communication. We need to understand alternative means of detecting and preventing algorithmic collusion such as transparency requirements, restrictions on certain types of algorithms, or real-time monitoring tools. 

\paragraph{Accountability}
Algorithmic accountability can play an important role in this context \citep{horneber2023algorithmic}. The concept refers to the responsibility of organizations and individuals to ensure that algorithms operate fairly, transparently, and ethically.  
In particular, transparency in algorithmic design could be an important tool in minimizing the risk of collusion.
Transparency assumes center stage in the Preventing Algorithmic Collusion Act of 2024,\footnote{\url{https://www.congress.gov/bill/118th-congress/senate-bill/3686/text}} a bill that has been introduced in the US Senate. Transparency is also an important aspect of the European Union's Digital Markets Act (DMA)\footnote{\url{https://digital-markets-act.ec.europa.eu/}} and the Digital Services Act (DSA),\footnote{\url{https://commission.europa.eu/strategy-and-policy/priorities-2019-2024/europe-fit-digital-age/digital-services-act_en}} although they do not yet address algorithmic collusion.  
By establishing design guidelines that discourage collusive strategies or by requiring algorithms to be transparent in their decision-making processes, firms may be able to mitigate unintended collusive behavior. Research might explore how different levels of transparency and design constraints affect algorithmic behavior and whether greater openness among algorithms would reduce or inadvertently increase the likelihood of collusion. Ultimately, the research in this field might lead to rules that can be implemented in regulatory audits and compliance tests. Firms might be asked to disclose their use of pricing algorithms and ensure that these tools are designed to comply with antitrust laws. Such measures aim to create an environment where algorithmic behaviors are subject to scrutiny, thereby discouraging collusive outcomes \citep{beneke2021remedies}. It is crucial to understand the regulatory measures needed to effectively minimize the risk of algorithmic collusion.

\paragraph{Beyond oligopoly competition}
While the focus of research on algorithmic collusion is on traditional oligpoloy models, there is no reason to believe that the phenomenon can only arise there. \citet{bichler2023low} analyzed display advertising auctions which are known to be automated via learning agents. Some recent research also explores algorithmic collusion in platform competition models such as that for ride-hailing or video streaming \citep{bichler2025platform}. Pricing of platforms on two-sided markets has drawn substantial attention in the BISE literature and the impact of algorithmic pricing describes a natural extension of this research \citep{dou2021platform, constantinides2018introduction, parker2016platform}. 

A central question in the economic sciences has long been under which conditions efficient outcomes can be achieved with market mechanisms. The welfare theorems provide conditions for a competitive equilibrium to exist that is Pareto efficient \citep{varian2014intermediate}. Over decades, the game-theoretical literature identified conditions under which efficient outcomes can be expected in equilibrium. Game theory highlights the role of incentives and strategic interaction in markets, and the Nash equilibrium assumes center stage. In game theory, agents are assumed to have enough information and that they can derive Nash equilibrium strategies that they use from the start. 

On real-world markets, agents often don't have the required information to derive an equilibrium, and even if they had, the equilibrium problem is computationally hard in general \citep{daskalakis2009complexity}. Importantly, agents don't have the information necessary about their competitors costs or values to derive an equilibrium. This is why on automated markets, agents rely on learning algorithms, exploring various prices, and exploiting the information gained over time. There is limited understanding of the circumstances under which such repeated interactions among such learning agents yield efficient outcomes. However, this understanding is important to understand the outcomes of algorithmic pricing in online retail markets. Gaining insights into these algorithmic markets represents an important and fundamental challenge.

The BISE community has long analyzed electronic markets. The event of learning algorithms in this market is more than a detail. It is fundamental that we understand how such learning agents impact the outcome of markets and whether such algorithmic markets can be expected to be efficient. This requires technical understanding of learning algorithms as well as the economic principles of market institutions.  As such, the Business \& Information Systems Engineering community is well equipped to address this challenging and yet hardly understood phenomenon. 

\section*{Acknowledgments}
This project was funded by the Deutsche Forschungsgemeinschaft (DFG, German Research Foundation) - GRK 2201/2 - Project Number 277991500 and BI 1057/9.



\bibliographystyle{informs2014} 


\end{document}